\documentstyle[aps,pre,epsf,twocolumn]{revtex}

\begin{document}
\draft
\title{Heat conduction in the diatomic Toda lattice revisited}
\author{Takahiro Hatano}
\address{Department of Pure and Applied Sciences, 
University of Tokyo, Komaba, Tokyo 153-0041, Japan}
\date{\today}
\maketitle

\begin{abstract}
The problem of the diverging thermal conductivity in one-dimensional 
(1-D) lattices is considered.
By numerical simulations, it is confirmed that 
the thermal conductivity of the diatomic Toda lattice diverges, 
which is opposite to what one has believed before.
Also the diverging exponent is found to be almost the same as 
the FPU chain.
It is reconfirmed that the diverging thermal conductivity is 
universal in 1-D systems where the total momentum preserves.
\end{abstract}

Heat conduction in one-dimensional lattice is rather an old problem.
Many authors have investigated the property of thermal conductivity 
to understand what ingredients are essential to the standing of 
a macroscopic law, that is, the Fourier law; 
\begin{equation}
\label{fourier}
\langle j\rangle=-\kappa\bigtriangledown T, 
\end{equation}
where $\kappa$ is a thermal conductivity.
It is well known that in integrable systems such as 
harmonic chains or ideal gas, the Fourier law is not 
valid since no temperature gradient is formed \cite{rieder}, 
while various numerical simulations of 
nonintegrable systems show temperature gradients.
However, it is also found that thermal conductivity of 
nonintegrable systems such as the FPU chain diverge as 
$N^{\alpha}$ \cite{machida,lepri1} where N is the degree of freedom.
In other words, thermal conductivity becomes infinite 
in the thermodynamic limit.

On the other hand, finite conductivities are seen in some 
1-D nonintegrable systems.
Casati {\em et al.} invented so-called ding-a-ling model 
consisting of alternate harmonic oscilators and free particles, 
and found that the model has the finite conductivity \cite{casati}.
The similar kind of the model which has finite conductivity 
is also investigated by Prosen and Robnik \cite{prosen}.
Most recently, Hu {\em et al.} found that the Frenkel-Kontrova model 
has the finite conductivity \cite{hu}.
In these models, the conductivity converges at the certain value 
with relatively small $N$ which does not exceed $100$.
This convergence makes apparent contrast with the FPU chain 
where the conductivity still grows even at $N\simeq 5000$ 
\cite{machida,lepri2}.
As the common feature of these systems which have finite 
conductivity, the external field is introduced to confine 
the movement of each particle.
The Hamiltonian of the systems are represented as 
\begin{equation}
\label{external}
H=\sum_i[\frac{p_i^2}{2m}+U(x_{i+1}-x_i)+V(x_i)], 
\end{equation}
where $V(x)$ is the external trapping potential.
At this point, one might think that the external field plays 
the key role for obtaining the finite condutivity \cite{hu}.
However, the finite condutivity is also obtained for 
the diatomic Toda lattice (DTL) \cite{jackson} 
whose Hamiltonian is written as 
\begin{equation}
\label{toda}
H=\sum_i[\frac{p_i^2}{2m_i}+\exp(x_{i+1}-x_i)], 
\end{equation}
where $m_i$ denotes the mass of alternate two different particles.
The DTL has no external potential which is different from 
Eq.(\ref{external}).
It has been still under cover what is responsible for 
the finite conductivity.

Recently, Lepri {\em et al.} found that the autocorrelation 
function of the total heat current vanishes like $t^{-0.6}$ 
in the FPU chain \cite{lepri3}.
This implies the divergence of the thermal conductivity 
as a result of the Green-Kubo formula; 
\begin{equation}
\label{greenkubo}
\kappa=\lim_{t\rightarrow\infty}\lim_{V\rightarrow\infty}
\frac{1}{Vk_BT^2}\int_0^t dt'\langle J(0)J(t')\rangle ,
\end{equation}
where $J(t)=\int j(x,t)dx$, and $V$ is the volume of the system.

Indeed, due to the conservation laws, long-time tails of 
the correlation functions are quite general results in fluids 
\cite{ernst}.
The rough explanation is as follows.
Hydrodynamically, the local heat current $j(x,t)$ is expressed as 
\begin{equation}
\label{localcurrent}
j(x,t)=h(x,t)v(x,t)-\kappa\bigtriangledown T(x,t), 
\end{equation}
where $h(x,t)$ and $v(x,t)$ denote local enthalpy density 
and local velocity of the fluid, respectively \cite{reichl}.
Since $v(x,t)$ appears in the first term of 
Eq.(\ref{localcurrent}), an autocorrelation function of 
the total heat current $\langle J(0)J(t)\rangle$ includes 
effect of the velocity autocorrelation function (VACF) 
\cite{pomeau}.
In the system where the total momentum preserves, 
asymptotic behavior of the VACF is propotional to $t^{-d/2}$, 
where $d$ is the dimensionality of the system.
Then $\langle J(0)J(t)\rangle$ also decays like $t^{-d/2}$, 
which implies the divergence of the integral of 
Eq. (\ref{greenkubo}) for $d\le 2$.
In the system where the total momentum does not preserve, 
the VACF vanishes much faster than that.
For example, in the Lorentz gas the VACF decays like 
$-t^{-d/2-1}$ \cite{hauge}.
In those systems the VACF does not cause the divergence of 
Eq. (\ref{greenkubo}).
We remark that the contribution of the second term of 
Eq. (\ref{localcurrent}) to $\langle J(0)J(t)\rangle$ is 
$t^{-d/2-1}$.
This term is not responsible for the diverging conductivity.

Those explanations account for the diverging conductivity 
in the FPU chain, and also the finite ones of the models where 
the total momentum does not preserve due to the external field, 
such as the ding-a-ling model.
However, the explanation does not apply to the diatomic 
Toda lattice where the total momentum preserves.
The fact that the DTL has a finite thermal conductivity 
has been invoking confusions.
In this Rapid Communication, 
we recheck the result of Ref. \cite{jackson} 
to find out what is really going on in the DTL.

The Hamiltonian of the DTL is given by Eq. (\ref{toda}).
We perform numerical simulations of the DTL in contact with two 
thermal reservoirs whose temperatures are denoted as $T_1$ and $T_2$.
Note that the choice of models for thermal reservoirs is critical, 
since there might exist the temperature gaps at the extrema of 
the lattice connecting with the reservoirs.
It makes the definition of temperature gradient ambiguous, 
because the system will not obey the assigned boundary conditions; 
i.e. temperatures of the thermal reservoirs.
Since thermal conductivity is defined as 
$\langle j\rangle/\bigtriangledown T$, 
it is important to determine $\bigtriangledown T$ 
exclusively by controll parameters.
The model we adopt here is the thermal wall type 
\cite{casati,thermalwall}.
When the particle collides with the wall, 
it reflects the particle back with a new momentum $p$ at random.
The probability distribution function of $p$ is given by 
\begin{equation}
\label{reservoir}
\phi (p)= \frac{|p|}{mk_BT}\exp[-\frac{p^2}{2mk_BT}].
\end{equation}
The local heat flux $j_l(t)$ is defined as the energy transfer per 
unit time from the l-th particle to the (l+1)-th particle.
\begin{equation}
\label{define_j}
j_l(t)=\frac{\partial U(x_l-x_{l+1})}{\partial x_l}v_l.
\end{equation}
The total heat current appearing in the Green-Kubo formula is 
\begin{equation}
\label{totalj}
J(t)=\sum_{l=1}^N j_l(t) a, 
\end{equation}
where $a$ is the average distance between two particles.
The average current is then defined as 
\begin{equation}
\label{define_av_j}
\langle j\rangle=\frac{1}{T}\int_0^Tdt \frac{1}{aN}J(t).
\end{equation}

Hereafter we fix the mass ratio of the particles to be $0.5$, 
that is, $m_{2n-1}=2m_{2n}$.
The temperatures of the thermal reservoirs are set to be $100$ and $10$.
Note that all these conditions are the same as Ref. \cite{jackson} 
except for the reservoir model.
Numerical integration is done by the symplectic integrator 
of the fourth-order \cite{yoshida} in order to preserve 
the symplectic structure of the phase space.
Note that the distance between two thermal walls is $aN$ 
so that the average density is fixed regardless of the number 
of particles.
We set $a=1$ and $m_{2n}=1$ for non-dimensionization.

First we check the temperature profile.
We define the temperature of the l-th site as the long time average 
of $m_lv_l^2$ based on the virial theorem.
The result is shown in Fig. 1.
Since no gap is seen at the extrema, 
temperature gradient $\bigtriangledown T$ becomes $(T_1-T_2)/N$.
We can safely define the thermal conductivity as 
\begin{equation}
\label{kappa}
\kappa =\frac{\langle j\rangle N}{T_1-T_2}, 
\end{equation}
where $\langle j\rangle$ is defined by Eq.(\ref{define_av_j}).
The system size dependence of the thermal conductivity 
is shown in Fig. 2.
It is clearly seen that the conductivity diverges like $N^{0.35}$.
The exponent $0.35$ is very close to the one for the FPU chain (0.38).
It is reasonable to consider that the origin of the divergence 
is the same as the case of the FPU chain, i.e. 
the long-time tail of the Green-Kubo integrand.
We check an autocorrelation function of the total heat current 
$\langle J(0)J(t)\rangle$, 
by taking a periodic boundary condition instead of thermal walls.
The initial condition is chosen within the microcanonical ensemble 
whose temperature is $(T_1+T_2)/2=55$.
Fig. 3 clearly shows the long-time tail which is 
approximately propotional to $t^{-0.65}$ just like the FPU chain.
This long-time tail is the strong evidence for the diverging 
thermal conductivity in Fig. 2, 
and also helps the unified understanding of 
the heat conduction in 1-D lattices.

However, one may think that the temperature difference adopted here 
is so large that the linear response theory does not apply.
To answer the suspicion, we check the thermal conductivity at 
the smaller temperature gradient which is closer to equilibrium, 
i.e. $T_1=5$, $T_2=4$.
The system at this temperatures also shows the divergence of 
$N^{0.35}$ and the long-time tail of $t^{-0.65}$.

In order to confirm the divergence in the diatomic Toda lattice, 
we also test other versions of the DTL \cite{toda}; 
\begin{eqnarray}
\label{hardsphere}
H&=&\sum_i\frac{p_i^2}{2m_i}+ hard core, \\
\label{bounded}
H&=&\sum_i[\frac{p_i^2}{2m_i}+\exp(x_i-x_{i+1})+x_{i+1}-x_i].
\end{eqnarray}
For thermal reservoirs, we use thermal wall model as before 
in the diatomic hard spheres of Eq. (\ref{hardsphere}).
Note that thermal reservoir employed in the simulation of 
Eq. (\ref{bounded}) is the Langevin type.
\begin{eqnarray}
\label{langevin1}
m_1\dot{v}_1+\zeta v_1+\xi_1 (t)&=&1-\exp(x_1-x_2), \\
\label{langevin2}
m_N\dot{v}_N+\zeta v_N+\xi_2 (t)&=&-1+\exp(x_{N-1}-x_N), 
\end{eqnarray}
where $\xi_i(t)$ denotes the Gaussian white noise.
($\langle\xi_i(t)\rangle=0$ and 
$\langle\xi_i(0)\xi_i(t)\rangle =2\zeta k_BT_i\delta (t)$.)
We set $T_1=5$ and $T_2=4$ for the both model.
System size dependences of the thermal conductivity of 
these models are shown in Fig. 2.
They also show the divergence of $N^{0.33}\sim N^{0.37}$.

The result obtained in this Rapid Communication is 
quite opposite to the results of Jackson {\em et al.} \cite{jackson}.
The keypoint is the formation of the temperature gradient.
In Ref. \cite{jackson}, the temperature profile has large gaps 
at the extrema of the lattice so that the real temperature gradient 
gets smaller than $N/(T_1-T_2)$.
Hence it is improper to define the thermal conductivity as 
$\langle j\rangle N/(T_1-T_2)$ as they did.
Moreover, since the size of the gap may depend on $N$, 
system size dependence of the thermal conductivity measured 
in that way is not precise.

The existence of the gaps is due to the model of the heat bath.
In Ref. \cite{jackson}, the new momenta are randomly given 
to the end particles of the lattice. 
Although the distribution function is the same as ours, 
i.e. Eq. (\ref{reservoir}), the new momenta are given 
at finite time steps which is determined randomly 
from the uniform distribution.
When the average time interval is shorter than the relaxation time 
of the lattice, the gap is formed.
This issue has been partially reported in Refs. \cite{prosen,lepri2}.
In our models, for instance, the Langevin model represented by 
Eqs. (\ref{langevin1}) and (\ref{langevin2}) yields temperature gaps 
when $\zeta$ becomes large.

In this Rapid Communication, 
we have confirmed that the thermal conductivity 
of the diatomic Toda lattice diverges as $N^{0.35}$ just like 
the FPU chain.
This divergence is gereric in the 1-D momentum preserving systems, 
due to the long-time tails in the Green-Kubo integrands.
Only are the systems where the total momentum does not preserve 
and the 3-D fluids expected to have the finite 
thermal condutivities in the thermodynamic limit.

However, it is still unclear that the quantitative conditions 
for the existence of temperature gradients, 
aside from the choice of the heat bath model.
Nonintegrability itself is the necessary condition.
Quantitative study of the transport processes 
from the viewpoint of dynamical systems must be 
the main focus of the future problem.

The author is grateful to S. Sasa for critical comments and 
encouragements.
The author also thanks T. Shibata, S. Takesue, K. Kaneko, M. Machida, 
and K. Saito for stimulating discussions.


\begin{figure}[htbp]
\label{fig1}
\caption{Temperature profile formed in the diatomic Toda lattice.
The temperatures of the reservoirs are 100 and 10. 
System size $N$ is $1000$. 
The shape of the profile will not change with the increase of $N$.}
\end{figure}

\begin{figure}[htbp]
\label{fig2}
\caption{System size dependence of the thermal conductivity.
Circles correspond to the diatomic Toda lattice 
with $T_1=100$, $T_2=10$.
Squares denote diatomic hard spheres of Eq. (11).
Triangles represent another version of the DTL written as Eq. (12).
The solid line is propotional to $N^{0.35}$.}
\end{figure}

\begin{figure}[htbp]
\label{fig3}
\caption{The autocorrelation function of the total heat current 
with the periodic boundary condition.
Dashed line is propotional to $t^{-0.65}$.
System size N is $2000$.}
\end{figure}

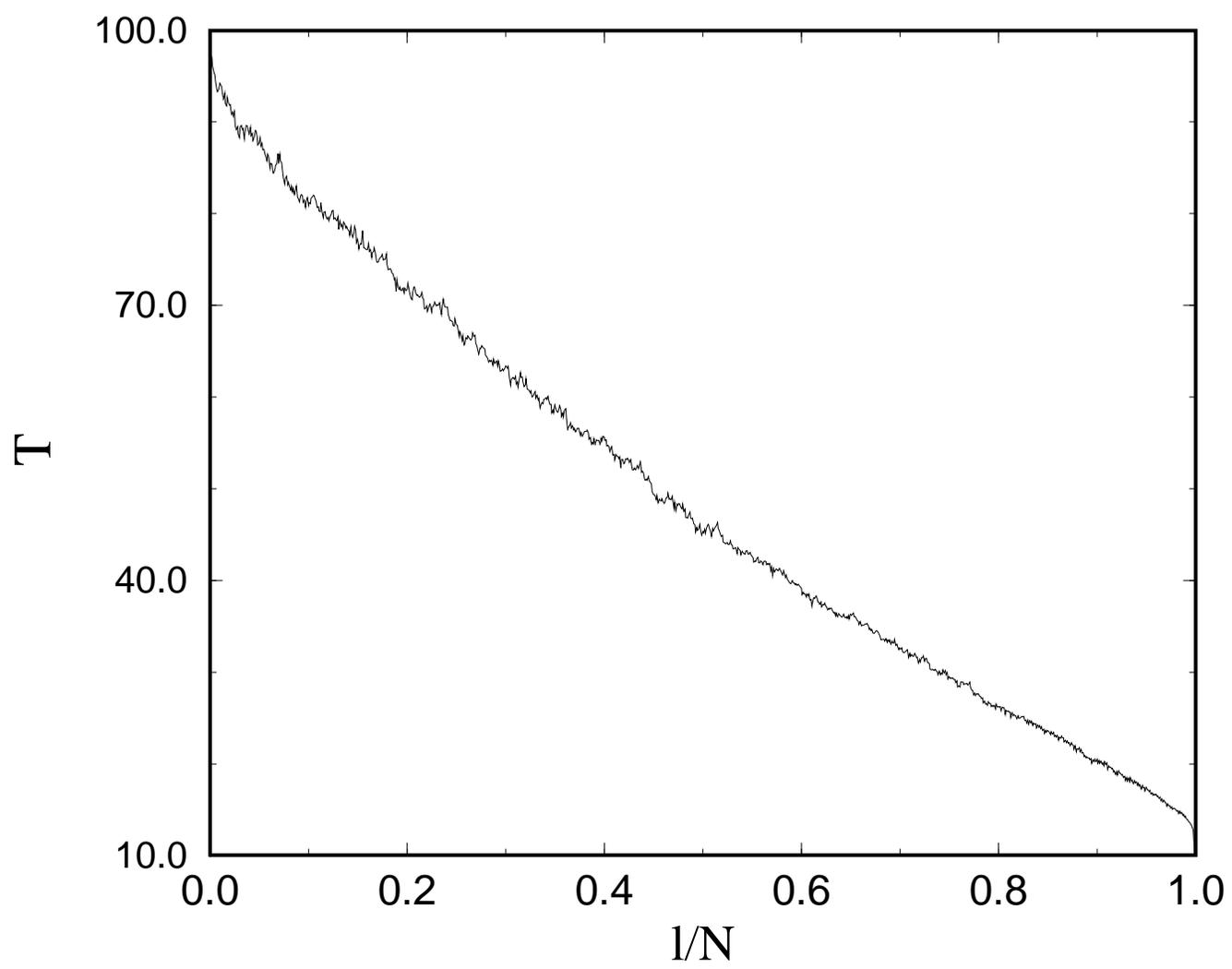
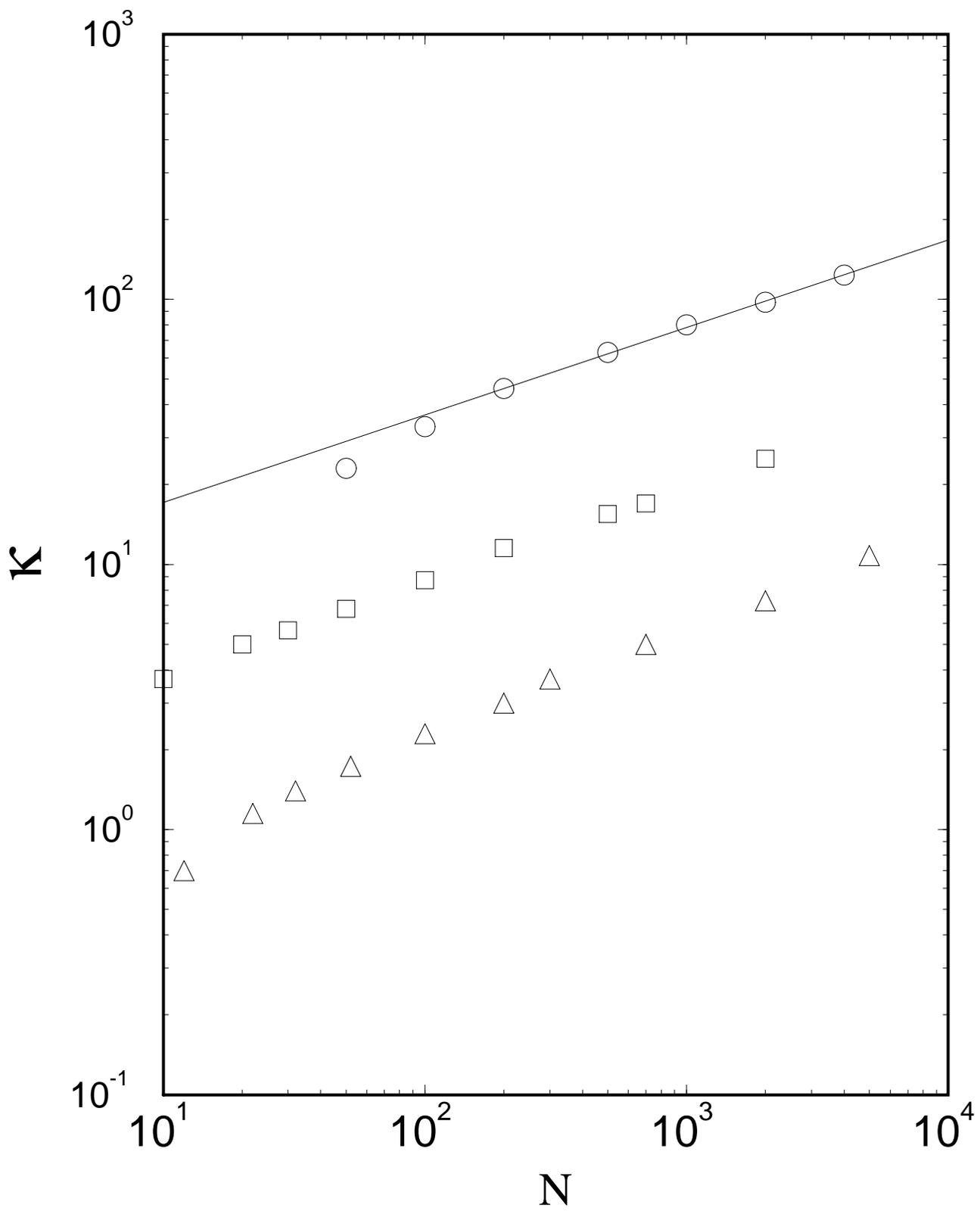
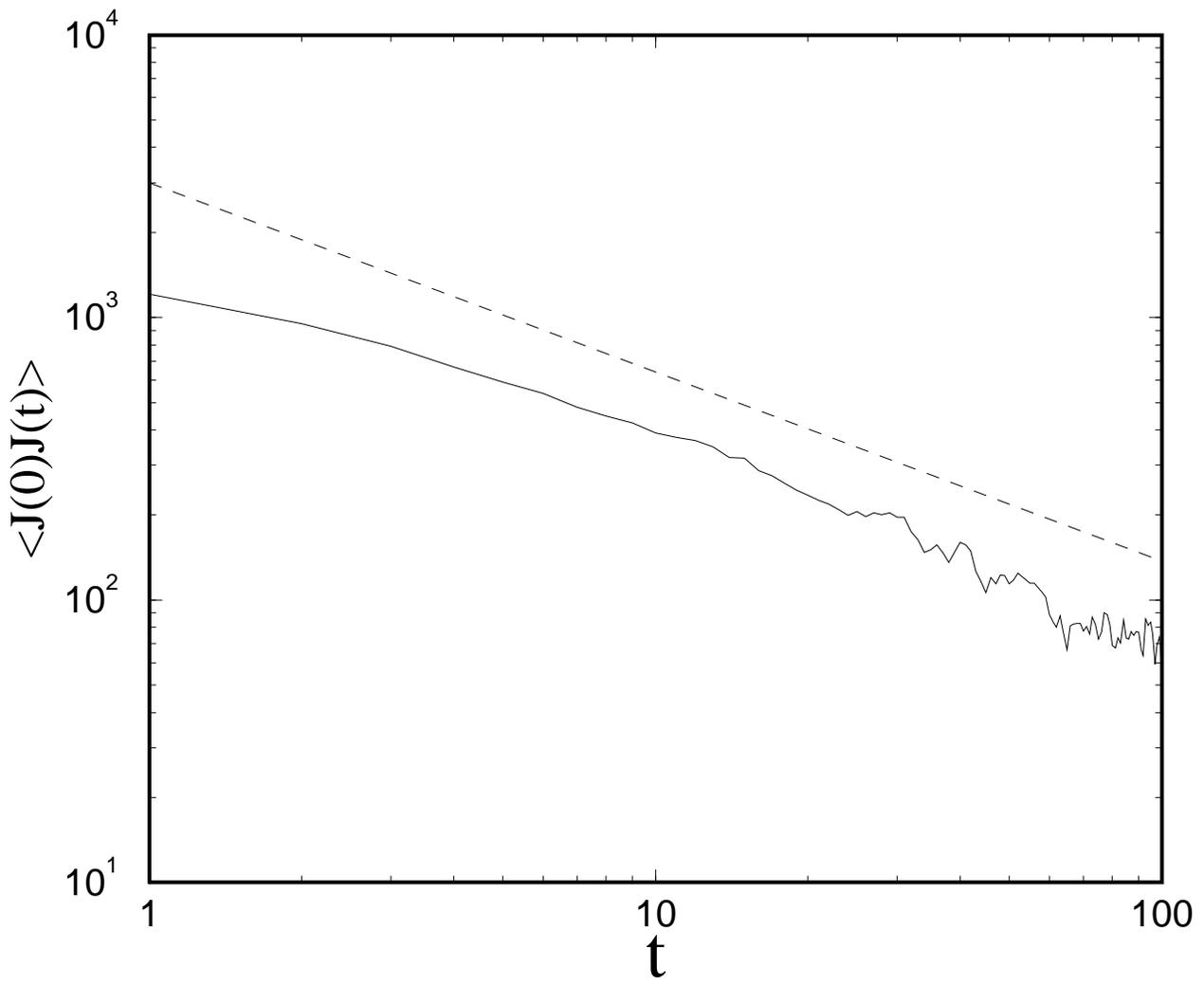
\end{document}